\documentclass[12pt]{article}

\newcommand{\be}{\begin{equation}}
\newcommand{\ee}{\end{equation}}
\newcommand{\bi}[1]{\vspace{-3mm} \bibitem{#1}}
\usepackage{epsfig,amsmath,amssymb,graphics,graphicx}

\begin{document}

\begin{center}
{\it Journal of Physics A 41 (2008) 035101}
\end{center}

\vskip 11 mm

\begin{center}
{\Large \bf Chains with Fractal Dispersion Law}
\vskip 7 mm

{\large \bf Vasily E. Tarasov }\\

\vskip 3mm

{\it Skobeltsyn Institute of Nuclear Physics, \\
Moscow State University, Moscow 119991, Russia } \\
E-mail: tarasov@theory.sinp.msu.ru

\vskip 3mm

\begin{abstract}
Chains with long-range interactions are considered.
The interactions are defined such that each
$n$th particle interacts only with chain particles 
with the numbers $n \pm a(m)$, where $m=1,2,3,...$ and 
$a(m)$ is an integer-valued function.
Exponential type functions $a(m)=b^m$, where $b=2,3,..$, are discussed. 
The correspondent pseudodifferential equations of chain oscillations
are obtained. Dispersion laws of the suggested chains are described by 
the Weierstrass and Weierstrass-Mandelbrot functions.
\end{abstract}

\end{center}


\vskip 11 mm

%%%05.45.Df Fractals
%%%45.05.+x General theory of classical mechanics of discrete systems
%%%45.50.-j Dynamics and kinematics of a particle and a system of particles

\section{Introduction}

Long-range interaction (LRI) has been the subject of investigations for a long time.
An infinite Ising chain with LRI was considered by Dyson \cite{Dyson}.
The $d$-dimensional Heisenberg model with long-range interaction 
is described in Refs. \cite{J,CMP}, and 
its quantum generalization can be found in \cite{NakTak,S}.  
Solitons in a chain with the long-range Lennard-Jones-type interaction 
were considered in \cite{Ish}. 
Kinks in the Frenkel-Kontorova chain model with long-range 
interparticle interactions were studied in \cite{BKZ}. 
The properties of time periodic spatially localized solutions (breathers) 
on discrete chains in the presence of algebraically decaying LRI 
were described in \cite{Br4}. 
Energy and decay properties of discrete breathers in chains with LRI 
have also been studied in the framework of the Klein-Gordon 
\cite{BK}, and discrete nonlinear Schrodinger equations \cite{Br6}. 
A main property of the chain dynamics 
with power-like long-range interactions \cite{Br4,AEL} is that 
the solutions of chain equations have power-like tails. 
The power-law LRI is also considered in Refs. \cite{LZ,TZ3,KZT,KZ}. 
In Ref. \cite{JPA}, we formulate the consistent definition of 
continuous limit for the chains with long-range interactions (LRI).
In the continuous limit, the chain equations with power-like LRI give  
the medium equations with fractional derivatives.

Usually we assume for LRI that each chain particle acts 
on all chain particles.
There are systems where this assumption cannot be used. 
In general, the chain cannot be considered as a straight line.
For example, the linear polymers can be represented as some compact objects.
It is well known that "tertiary structure" of proteins refers 
to the overall folding of the entire polypeptide chain 
into a specific 3D shape \cite{Holde,PDB,KS}. 
The tertiary structure of enzymes 
is often compact, globular shaped \cite{Holde,PDB}.
In this case, we can consider that the chain particle is interacted with
particles of a ball with radius $R$.
Then only some subset $A_n$ of chain particles act on $n$th particle. 
We suppose that $n$th particle is interacted only with $k$th particles with 
$k=n \pm a(m)$, where $a(m) \in \mathbb{N}$ and $m=1,2,3,...$.
We can consider fractal compactified linear polymers (chains), such that
these "compact objects" satisfy the power law $N(R) \sim R^{D}$, 
where $2<D<3$ and $N(R)$ is a number of chain particles 
inside the sphere with radius $R$.
As an example of such case, $a(m)$ will be described by 
exponential type functions $a(m)=b^m$, 
where $b>1$ and $b \in \mathbb{N}$.
In this case, the LRI will be called fractal interaction.

The goal of this paper is to study a connection between  
the dynamics of chain with fractal long-range interactions (FLRI)
and the continuous medium equations with fractal dispersion law. 
Here, we consider the chain of coupled linear oscillators with FLRI. 
We make the transformation to the continuous field 
and derive the continuous equation which describes 
the dynamics of the oscillatory medium.
We show how the oscillations of chains  
with FLRI are described by the fractal dispersion law.
This law is represented by the Weierstrass functions 
whose graphs have non-integer box-counting dimension, i.e., 
these graphs are fractals. 
Fractals are good models of phenomena and objects 
in various areas of science \cite{Mand}. 
Note that fractals in quantum theory  have recently 
been considered in \cite{Kogen}. 
In this paper, we prove that the chains with long-range interaction
can demonstrate fractal properties described by fractal functions.

%%%%%%%%%%%%%%%%%%%%%%%%%%%%%%%%%%%%%%%%%%%%%%%%%%%%%%%%%%%%%%%%%%%%%%%%%%
\section{Chain equation}

One of the oldest fractal functions is Weierstrass function \cite{Weier}:
\be \label{Wx} 
W(x) =\sum^{\infty}_{n=0} b^n \, \cos (a^n \pi x) \ee
introduced as an example of everywhere continuous 
nowhere differentiable function by Karl Weierstrass around 1872. 
Maximum range of parameters for which
the above sum has fractal properties was found by 
Godfrey Harold Hardy \cite{Hardy} in 1916, who showed that
\[ 0<b<1, \quad ab \ge 1 . \]
The box-counting dimension of the graph of the Weierstrass function $W(x)$ is
\be \label{DW}
D=2+\frac{\ln(b)}{\ln(a)}=2-\Bigl|\frac{\ln(b)}{\ln(a)}  \Bigr| . \ee
Functions whose graphs have non-integer box-counting dimension 
are called fractal functions. 

Consider a one-dimensional system of interacting oscillators that 
are described by the equations of motion,
\be \label{eq1}
\frac{d^2}{dt^2} u_n(t) =
c^2 \sum^{+\infty}_{m=-\infty} \frac{b(m)}{h^2} \, 
[u_{n+a(m)}(t)-2u_n(t)+u_{n-a(m)}(t)] ,  \ee
where $u_n(t)$ are displacements from the equilibrium, and
$h$ is the distance between the oscillators. 
Here $a(m)$ and $b(m)$ are some functions of integer number $m$.
The case $b(m)=J(|n-m|)$ has been considered in \cite{JPA}.
The right-hand side of the equation
describes an interaction of the oscillators in the system.

We illustrate this chain equation with well-known example \cite{Maslov}.
In the case of nearest-neighbor interaction, we have
\[ a(m)=1, \quad b(m)= \delta_{m0} . \]
Then equation (\ref{eq1}) gives
\be \label{eq2}
\frac{d^2}{dt^2} u_n(t) =
\frac{c^2}{h^2} \, [u_{n+1}(t)-2u_n(t)+u_{n-1}(t)]  .  \ee
We can define a smooth function $u(x,t)$ such that
\be \label{eq3} u(nh,t)=u_n(t) . \ee
Then equation (\ref{eq2}) has the form
\be \label{eq4}
\partial^2_t u(x,t)=\frac{c^2}{h^2} \, [u(x+h,t)-2u(x,t)+u(x-h,t)] . \ee
This is the differential-difference equation.
Using the relation
\[ \exp(-ih\partial_x) \, u(x,t)=u(x+h,t) , \]
we can rewrite equation (\ref{eq4}) in the form
\be \label{eq6}
 \partial^2_t u(x,t)+\frac{4c^2}{h^2} \, 
\sin^2 \Bigl(-\frac{ih}{2} \partial_x\Bigr) \, u(x,t)=0 . \ee
This is the pseudo-differential equation.
The properties of this equation have been considered in \cite{Maslov}.
For $h \ \rightarrow \ 0$, we obtain the wave equation
\[ \partial^2_t u(x,t)-c^2\partial^2_x u(x,t)=0 . \]

%%%%%%%%%%%%%%%%%%%%%%%%%%%%%%%%%%%%%%%%%%%%%%%%%%%%%%%%%%%%%%%%%%%%%%%%%%
\section{Fractal long-range interaction}

If $a(m)$ in equation (\ref{Wx}) is not a constant function, 
then we have the long-range interaction of the chain particles. 
Note that the function $a(m)$ should be integer-valued.
For example, $a(m)=m$ and $a(m)=2^m$. 
The set $A_n=\{n \pm a(m): \ m \in \mathbb{N}\}$ 
describes the numbers of particles
that act on the $n$th particle.\\

(a) If $a(m)=m$, where $m \in \mathbb{N}$, then
$A_n$ is a set of all integer numbers $\mathbb{Z}$ for all $n$, 
i.e., $A_n=\mathbb{Z}$.
In this case, the $n$th particle interacts with all chain particles. \\

(b) If $a(m)=2^m$, where $m \in \mathbb{N}$, then
$A_n$ is a subset of $\mathbb{Z}$, i.e., $A_n \subset \mathbb{Z}$.
In this case, the $n$th particle interacts only with chain particles
with numbers $n \pm 2$, $n \pm 4$, $n \pm 8$, $n \pm 16$ .... \\

The power law $a(m)=b^m$, where $b \in \mathbb{N}$, and $b>1$, 
can be realized for compact structure of linear polymer molecules.
For example, a linear polymer molecule is not a straight line.
Usually this molecule can be considered as a compact object. 
It is well-known that "tertiary structure" of proteins refers 
to the overall folding of the entire polypeptide chain 
into a specific 3D shape \cite{Holde,PDB,KS}. 
The tertiary structure of enzymes is often a compact, 
globular shape \cite{Holde,PDB}.
In this case, we can consider that the chain particle is interacted with
particles inside a sphere with radius $R$.
Then only some subset of chain particles act on $n$th particle.
We assume that  $n$th particle is interacted only with $k$th particles with 
$k=n \pm a(m)$, where $a(m) \in \mathbb{N}$ and $m=1,2,3,...$.
The polymer can be a mass fractal object \cite{Nano}. 
For fractal compactified linear polymer chains, 
we have the power-law $N(R) \sim R^{D}$, where $2<D<3$ and $N(R)$ 
is the number of chain particles in the ball with radius $R$.
Then we suppose that $a(m)$ is exponential type function such that $a(m)=b^m$, 
where $b>1$ and $b \in \mathbb{N}$. This function
defines the fractal long-range interaction. \\

Let us consider equation (\ref{eq1}) 
with a fractal long-range interaction.
Using the smooth function (\ref{eq3}), we obtain the
differential-difference equation
\be \label{eq7}
\partial^2_t u(x,t)=\frac{c^2}{h^2} \sum^{+\infty}_{m=-\infty} b(m) \, 
[u(x+a(m)h,t)-2u(x,t)+u(x-a(m)h,t)] . \ee
Using 
\be \label{eq8}
\exp \{-i a(m)h\partial_x \} \, u(x,t)=u(x+a(m)h,t) ,
\ee
equation (\ref{eq7}) can be presented as the pseudo-differential equation
\be \label{eq9}
 \partial^2_t u(x,t)+\frac{4 c^2}{h^2} \sum^{+\infty}_{m=-\infty} b(m) \, 
\sin^2 \Bigl(-\frac{ih a(m)}{2} \partial_x\Bigr) \, u(x,t)=0 . \ee
For $a(m)=1$ and $b(m)=\delta_{m0}$, 
this equation gives equation (\ref{eq6}) that describes oscillations for
the case of the nearest-neighbor interaction.

Let us consider the pseudo-differential operator
\be \label{WMO}
{\cal L}=2 \sum^{+\infty}_{m=-\infty} b(m) \, 
\sin^2 \Bigl(-\frac{ih a(m)}{2} \partial_x\Bigr) . \ee
The function
\be \label{Psi}
\Psi(x,k)=A\exp (ikx) \ee
is an eigenfunction of this operator:
\[ {\cal L}\Psi(x,k)=\lambda(k) \Psi(x,k) . \]
Here $\lambda(k)$ is the eigenvalue of the operator, such that
\[ \lambda(k)=2 \sum^{+\infty}_{m=-\infty} b(m) \, 
\sin^2 \Bigl(\frac{h a(m)}{2} k \Bigr) . \]
Using $2\sin^2(\alpha/2)=1-\cos(\alpha)$, we obtain 
\be \label{lambda}
\lambda(k)= \sum^{+\infty}_{m=-\infty} b(m) \, [1- \cos (h a(m) k) ] . \ee
If 
\be \label{abm} a(m)=a^m , \quad b(m)=a^{(D-2)m} , 
\quad a \in \mathbb{N}, \quad a>1 , \ee
then equation (\ref{lambda}) has the form
\[ \lambda(k)= C(hk) , \]
where $C(z)$ is the cosine Weierstrass-Mandelbrot 
function \cite{Mand,Feder},
\[ C(z)= \sum^{+\infty}_{m=-\infty} a^{(D-2)m}  \, 
[1- \cos ( a^{(D-2)m} z)] .  \]
The box-counting dimension of the graph of this function is $D$. 
The operator (\ref{WMO}) can be called the Weierstrass-Mandelbrot operator.
The spectral graph $(k,C(hk))$ of this operator 
is a fractal set with dimension $D$.

Using the operator (\ref{WMO}), 
equation (\ref{eq9}) takes the form
\[ \partial^2_t u(x,t)+\frac{4 c^2}{h^2}{\cal L} u(x,t)=0 . \]
It is not hard to prove that the dispersion law for the chain 
with the long-range interactions is described by equations  
(\ref{eq1}) and (\ref{abm}) has the form
\[ \omega^2+\frac{2c^2}{h^2}C(hk)=0 . \]
Then the graph $(k,\omega(k))$ is a fractal.
Note that the group velocity 
$v_{group}={\partial \omega(k)}/{ \partial k}$
for the plane waves cannot be find, 
since $C(z)$ is the nowhere differentiable function. \\

Let us consider a generalization of conditions (\ref{abm}) in the form
\be \label{abm2} a(m)=a^m, \quad b(m)=b^m , \ee
where $a \in \mathbb{N}$, $b<1$, and $a>1$. 
For $b=a^{D-2}$, we have (\ref{abm}).
If we use (\ref{abm2}), we can obtain the pseudo-differential equation
\be \label{eq10}
\partial^2_t u(x,t)+\frac{4c^2}{h^2} \, 
\sin^2 \Bigl(-\frac{ih}{2} \partial_x\Bigr)  \, u(x,t)
+ \frac{8c^2}{h^2} \sum^{+\infty}_{m=1} b^m \, 
\sin^2 \Bigl(- \frac{ih a^m}{2} \partial_x \Bigr) \, u(x,t) . \ee
This equation can be presented in the form
\be \label{eq10b}
\partial^2_t u(x,t)+\frac{4c^2}{h^2} \, 
\sin^2 \Bigl(-\frac{ih}{2} \partial_x\Bigr) \, u(x,t)
+ M^2 c^2 u(x,t)=
4 \frac{c^2}{h^2} \sum^{+\infty}_{m=1} b^m \, 
\cos \Bigl(- ih a^m \partial_x \Bigr)  \, u(x,t) , \ee
where $M^2={4b}/{h^2(1-b)}$. 
Equation (\ref{eq10b}) describes the oscillations in the case 
of the fractal LRI.
The left-hand side of equation (\ref{eq10b}) 
in the limit $h \ \rightarrow \ 0$ gives the Klein-Gordon equation
\[ c^{-2}\partial^2_t u(x,t)-\partial^2_x u(x,t)+M^2 u(x,t)=0 .\]
The right-hand side of equation (\ref{eq10b}) describes 
a nonlocal part of the interaction.
Note that the pseudo-differential operator
\be \label{WO}
\Lambda= \sum^{+\infty}_{m=0} b^m \, \cos \Bigl(- ih a^m \partial_x \Bigr) , \ee
which is used in equation (\ref{eq10b}), 
has the eigenfunctions (\ref{Psi}) such that the eigenvalues is
the Weierstrass function $W(hk/\pi)$, where
$W(x)$ is defined by equation (\ref{Wx}). 
The box-counting dimension of the graph of 
the Weierstrass function $W(x)$ is (\ref{DW}).
The operator (\ref{WO}) can be called the Weierstrass operator.
The spectral graph $(k,W(hk/\pi))$ of this operator is a fractal set 
with dimension $2+\ln(b)/\ln(a)$.

%%%%%%%%%%%%%%%%%%%%%%%%%%%%%%%%%%%%%%%%%%%%%%%%%%%%%%%%%%%%%%%%%%%%%%%%%%
\section{Conclusion}

In this paper, we prove that the chains with long-range interaction 
can demonstrate fractal properties.
We consider chains with long-range interactions such that 
each $n$th particle is interacted only with chain particles 
with the numbers $n \pm a(m)$, where $m=1,2,3,...$. 
The exponential type functions $a(m)=b^m$, 
where $b>1$ is integer, are used to define 
a fractal long-range interaction. 
The equations of chain oscillations are characterized  
by dispersion laws that  are represented by Weierstrass and 
Weierstrass-Mandelbrot (fractal) functions.
The suggested chains with long-range interactions 
can be considered as a simple model for linear polymers 
that are compact, fractal globular shape.

%%%\newpage
%%%%%%%%%%%%%%%%%%%%%%%%%%%%%%%%%%%%%%%%%%%%%%%%%%%%%%%%%%%%%%%%%%%%%%%%%%

\end{document}